\documentclass[preprint,12pt]{elsarticle}

\usepackage{svg}

\usepackage{amssymb}
\usepackage{amsmath}
\usepackage[capitalize]{cleveref}
\usepackage{url}

\usepackage{subcaption}
\usepackage{hyphenat}
\journal{Acta Astronautica}

\begin{document}

\begin{frontmatter}



\title{Memristor-Based Neural Network Accelerators for Space Applications: Enhancing Performance with Temporal Averaging and SIRENs}


\author[label1,label2]{Zacharia A. Rudge\corref{cor1}} 
\author[label3]{Dominik Dold} 
\author[label1]{Moritz Fieback} 
\author[label2]{Dario Izzo} 
\author[label1]{Said Hamdioui} 

\cortext[cor1]{Corresponding author: z.a.rudge@tudelft.nl}

\affiliation[label1]{organization={TU Delft},
            addressline={Mekelweg 5}, 
            city={Delft},
            country={The Netherlands}}
            
\affiliation[label2]{organization={European Space Agency},
            addressline={Keplerlaan 1}, 
            city={Noordwijk},
            country={The Netherlands}}

\affiliation[label3]{organization={University of Vienna},
            addressline={Kolingasse 14-16}, 
            city={1090 Wien},
            country={Austria}}

\begin{abstract}
Memristors are an emerging technology that enables artificial intelligence (AI) accelerators with high energy efficiency and radiation robustness -- properties that are vital for the deployment of AI on-board spacecraft.
However, space applications require reliable and precise computations, while memristive devices suffer from non-idealities, such as device variability, conductance drifts, and device faults.
Thus, porting neural networks (NNs) to memristive devices often faces the challenge of severe performance degradation.
In this work, we show in simulations that memristor-based NNs achieve competitive performance levels on on-board tasks, such as navigation \& control and geodesy of asteroids.
Through bit-slicing, temporal averaging of NN layers, and periodic activation functions, we improve initial results from around $0.07$ to $0.01$ and $0.3$ to $0.007$ for both tasks using RRAM devices, coming close to state-of-the-art levels ($0.003-0.005$ and $0.003$, respectively).
Our results demonstrate the potential of memristors for on-board space applications, and we are convinced that future technology and NN improvements will further close the performance gap to fully unlock the benefits of memristors.
\end{abstract}




\end{frontmatter}

\section{Introduction} \label{sec:intro}
Artificial Intelligence (AI) -- in particular deep learning -- holds the power to revolutionize space missions by vastly increasing the autonomy of deployed spacecraft \cite{furano_towards_2020,kothari_final_2020, izzo_survey_2019,izzo_selected_2023,izzo2022neuromorphic}. Recent studies have demonstrated the practical application of deep neural networks in enhancing spacecraft autonomy, including uses such as real-time optimal control for asteroid landing \cite{cheng_real-time_2020}, trajectory generation and optimization \cite{ma_neural_2025} and minimum-fuel lunar landing \cite{ni_accelerating_2023}. However, deep learning is plagued by rapidly increasing energy requirements \cite{luccioni_power_2024} incompatible with most spaceborne applications.
Operating AI models in space imposes not only tight constraints on on-board energy consumption, but also strict requirements in terms of accuracy, area, temperature extremes, and robustness, e.g., radiation tolerance \cite{kothari_final_2020,schumann_radiation_2022}. 

Existing AI accelerators fail to fully address these needs. Most deep learning implementations are currently based on conventional GPU and CPU design paradigms, which generally do not meet the requirements for on-board application due to -- among other reasons -- the \textit{Von Neumann bottleneck} \cite{noauthor_beyond_2020}, i.e., the energy cost overhead of repeatedly moving data and instructions between the memory and the computational units, as well as technology scaling issues and their effect on radiation resilience \cite{amrouch_towards_2021}. 
These challenges underscore the necessity to explore beyond conventional architectures and device technologies in order to develop hardware-software solutions that enable us to fully unlock the potential of AI in space -- from low Earth orbit to deep space missions where power, communication bandwidths, and robustness requirements are pushed to the extreme.
A promising technology that has recently attracted attention in the space community is neuromorphic hardware \cite{izzo2022neuromorphic,lagunas2024performance,schumann_radiation_2022,lunghi2024investigation,arnold2025scalable,kucik_investigating_2021}: transistor-based chip designs that mimic the architecture of the brain to circumvent the Von Neumann bottleneck \cite{frenkel2023bottom}.
However, emerging digital neuromorphic accelerators \cite{schuman_survey_2017}, including Intel's Loihi \cite{davies_loihi_2018} and FPGA-based implementations, do not meet all space-specific requirements for similar reasons as conventional hardware accelerators \cite{montealegre_-flight_2015, schumann_radiation_2022,lunghi2024investigation,kucik_investigating_2021}. 
Similarly, analog neuromorphic accelerators \cite{izzo2022neuromorphic,arnold2025scalable,lunghi2024investigation}, while promising in terms of energy consumption, currently lack the performance guarantees and scalability required for space applications.

To solve the aforementioned issues and enable AI solutions that satisfy all requirements of on-board application, we propose adopting a new chip architecture based on one of the most promising emerging technologies for edge computing: in-memory computing using memristors \cite{singh_low-power_2021}. 
By integrating storage and computation into a single element,  chip designs based on memristors significantly reduce data movement \cite{diware_accurate_2023}, thereby alleviating the Von Neumann bottleneck. Neural network accelerators utilizing memristors, like PCM (Phase-Change Memory), RRAM (Resistive Random-Access Memory, also known as ReRAM), and MRAM (Magnetic Random-Access Memory) have demonstrated their ability to reduce energy consumption by orders-of-magnitude relative to conventional computing architectures \cite{khaddam-aljameh_hermes_2021, wan_compute--memory_2022, deaville_22nm_2022, bonnet_bringing_2023}. Memristors are non-volatile and do not exhibit leakage currents, but also enable highly parallel computing whilst being resilient to radiation \cite{hughart_comparison_2013}. RRAM devices in particular have been demonstrated to be radiation resistant to radiation events such as single event upsets during space flight \cite{lyu_research_2021}. Despite these benefits, the use of memristors also poses challenges, including the introduction of variability and non-ideal behaviors (e.g. read/write noise) that can strongly affect task performance -- oftentimes degrading it to levels far below those required in on-board applications \cite{rudge_guidance_2024}.

In this work, we close this gap through a simulation-based approach that captures the most crucial characteristics of memristive devices while allowing the evaluation of large-scale, state-of-the-art network architectures and on-board AI applications.
We demonstrate for two challenging on-board tasks that memristor-based SIREN neural networks equipped with layerwise temporal averaging reach competitive results -- despite the degrading effects of device non-idealities and analog noise. 
Although a reality gap remains both in terms of modeled device behaviors and the conditions encountered in space, we provide a faithful evaluation of memristive devices for realistic on-board applications.
More specifically, the contributions of this paper are as follows:
\begin{itemize}
    \itemsep0em
    \item \textbf{Implementation of memristor-based SIREN neural networks for realistic on-board AI applications.} We deploy and validate guidance and control networks (G\&CNETs) \cite{origer_guidance_2024} and neural networks for asteroid geodesy (geodesyNets ) \cite{izzo_geodesy_2022} on simulated PCM/RRAM hardware using the IBM Analog Hardware Acceleration Kit \cite{le_gallo_using_2023}.
    Compared to previous results \cite{rudge_guidance_2024}, we show that neural networks using periodic activation functions (SIRENs \cite{sitzmann_implicit_2020}) perform better on simulated memristive devices than their counterparts using ReLU-like activations, with G\&CNETs improving from a test loss of around $0.07$ to $0.02$.
    \item \textbf{Analysis of device non-idealities in space-relevant contexts.} We quantify the impact of device degradation and non-idealities (conductance drift, read/write noise, faulty devices) on task performance for both applications. In particular, we find that read/write noise strongly affects performance levels, while effects such as faulty devices are automatically compensated for during training in the case of geodesyNets.
    In contrast, G\&CNETs with periodic activation functions showed strong sensitivity to faulty devices and conductance drift despite good initial performance, highlighting the need for identifying neural networks that are robust to parameter perturbations after training. 
    \item \textbf{Recovering competitive performance via mitigation strategies.} We show that through bit-slicing and layerwise temporal averaging of neuronal outputs, competitive performance levels can be restored in both on-board applications -- with the test loss improving from around $0.02$ to $0.01$ for G\&CNETs and $0.3$ to $0.007$ for geodesyNets, getting close to competitive results (around $0.003 - 0.005$ and $0.003$, respectively).
    To our knowledge, this is the first work that demonstrates the feasibility of memristor-based accelerators for on-board space application tasks, showing that it is possible to reach performance levels satisfying mission requirements with this technology.
\end{itemize}

The remainder of this paper is structured as follows: \cref{sec:methods} details the experimental setup, including the architecture of G\&CNETs and geodesyNets, the simulation framework for the memristors and its non-idealities (e.g., conductance drift, read/write noise), and the implementation of non-ideality mitigation strategies such as temporal averaging and linear bit-slicing. \cref{sec:results} presents the performance of both G\&CNETs and geodesyNets, quantifies the impact of device non-idealities on each network, and evaluates the effectiveness of noise suppression techniques across both architectures. Finally, \cref{sec:discussion} discusses the significance of the results, limitations of the study, and future directions for memristor-based accelerators in on-board AI for spacecraft autonomy. 

\section{Methods} \label{sec:methods}

\subsection{On-board applications and models}

To evaluate the suitability of a memristor-based neural network accelerator, we first identify two space applications with classical neural network solutions that reach performance levels suitable for deployment on-board an actual spacecraft: guidance and control of a spacecraft (using G\&CNETs) and asteroid geodesy (using geodesyNets), i.e., estimating an asteroid's shape and mass distribution from sensory data.
These particular neural networks share similar light-weight architectures (feedforward with 100s of neurons and several layers), potentially allowing near-term prototyping with actual memristive crossbar arrays.
Moreover, both tasks require models producing high precision outputs, making them ideal candidates for testing the capabilities of memristor-based neural networks.
Finally, despite their architectural similarities, both tasks are quite different in nature: G\&CNETs solve a control task where the network output is applied to change the spacecraft's state, which in turn changes the input to the neural network.
In contrast, geodesyNets solve an inverse problem which involves evaluating an integral of the neural network over a given volume.
In the following, we describe both applications and neural network architectures in more detail.

\subsubsection{On-board guidance and control: G\&CNETs}
\begin{figure}[t]
    \centering
    \includegraphics[width=.95\columnwidth]{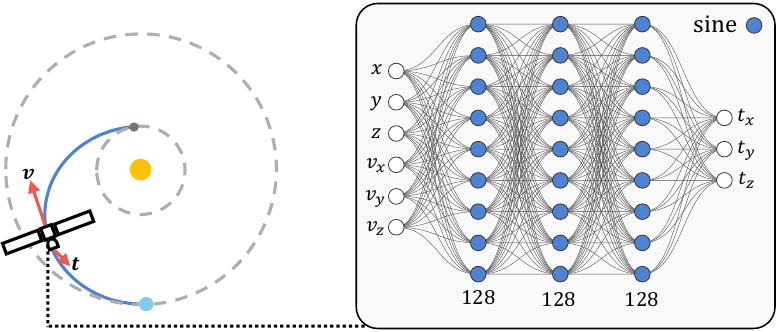}
	\caption{\textbf{(left)} A spacecraft transfers between two orbits using a G\&CNET. \textbf{(right)} The architecture of the G\&CNET studied. The network is composed of 3 hidden layers with 128 neurons each, using sine activation functions. It receives the spacecraft state (coordinates and velocity) as input and returns the thrust control.}
	\label{fig:gecnet_arch}
\end{figure}

Guidance and control networks, abbreviated as G\&CNETs, are fully-connected feedforward neural networks that take the state of a spacecraft as input and return an action, e.g., the thrust to control the spacecraft. 
Such networks are trained on a database of optimal state-action pairs, which are pre-computed by solving the two-point boundary value problem from different initial conditions $x_0$ resulting from the use of Pontryagin's maximum principle \cite{pontryagin_mathematical_1987}.
After training, the neural network represents the optimal state-feedback: A thrust control vector $\pmb t$, able to steer in (optimal) time $t^*_f$ a system from $x_0$ to a target state $x_f$, minimizing the cost function $\mathbf{J}(\mathbf{x},\mathbf{t},t_f) = \int_{t_0}^{t_f}\ell(\mathbf{x},\mathbf{t})dt$, where $\ell(x, t)$ denotes the instantaneous cost.
The resulting system serves as a suitable on-board substitute for more classical guidance and control in a variety of space applications, such as interplanetary transfer and asteroid landing. In the context of our work, we specifically aim to generate and track an interplanetary trajectory on-board, where the output control vector $\pmb t$ reflects the thrust action of a low-thrust spacecraft (\cref{fig:gecnet_arch}, left). 

For the presented experiments, we use a G\&CNET network with three hidden layers, each with 128 neurons (\cref{fig:gecnet_arch}, right). 
Our previous work on this topic evaluated the original architecture of the G\&CNETs with softplus and tanh activation functions \cite{izzo_interplanetary_2019}. However, recent advances have revealed that the performance of G\&CNETs is improved by employing the SIREN architecture \cite{origer_guidance_2024}.
Thus, we study here G\&CNETs with periodic activation functions, in our case a sine function $\sin(\omega_0\cdot x)$ with $\omega_0=1$ and input $x$. The weights of each layer are randomly initialized from a uniform distribution: for the first layer, we use $\mathcal{U}\big(-\frac{1}{n_i}, \frac{1}{n_i}\big)$, while for all other layers we $\mathcal{U}\big(-\frac{\sqrt{6/n_i}}{\omega_0}, \frac{\sqrt{6/n_i}}{\omega_0}\big)$ is used, with $n_i$ being the number of inputs of layer $i$.
Training is performed for 300 epochs.
Alternative methods and deep neural networks exist to perform similar tasks, such as Cheng et al.'s work on real-time optimal control for irregular asteroid landing \cite{cheng_real-time_2020}. In this case, G\&CNETs are surveyed as a representative example of a guidance and control deep neural network to validate the performance of memristor-based neural network accelerators.



\subsubsection{On-board geodesy of small bodies: geodesyNets}
\begin{figure}[t]
    \centering
    \includegraphics[width=.95\columnwidth]{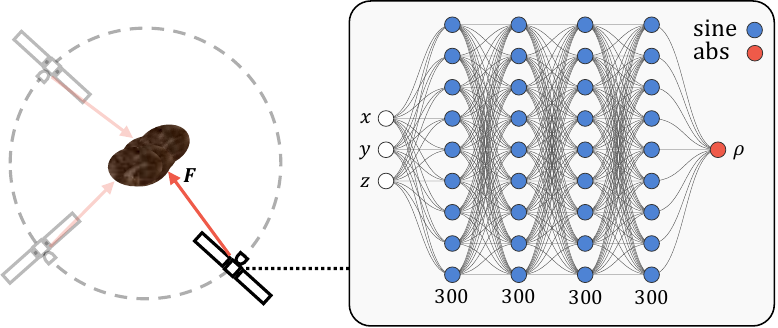}
	\caption{\textbf{(left)} A spacecraft orbiting a small, irregularly shaped body. Using the gravitational force (red), an on-board neural network (geodesyNet) learns to reconstruct the density field of the small body. \textbf{(right)} The architecture of the geodesyNet studied. Similar to the G\&CNET, the model is composed of several hidden layers with sine activations. It receives $x, y, z$ coordinates as inputs and returns the density of the asteroid at this coordinate. To calculate forces, an integral of the neural network has to be performed.}
	\label{fig:geodesy_arch}
\end{figure}

Geodesy of small bodies such as asteroids and comets is of great interest in a wide range of fields and industries, from purely scientific interest in the origins of our solar system \cite{glassmeier_rosetta_2007} to the commercial prospect of asteroid mining \cite{hein_techno-economic_2020}. Recently, a deep learning-based method for performing geodesy on-board a spacecraft using orbital acceleration measurements has been introduced: geodesyNets \cite{izzo_geodesy_2022}. 
GeodesyNets are inspired by Generative Query Networks \cite{eslami_neural_2018} and Neural Radiance Fields \cite{mildenhall_nerf_2021}, which introduce novel neural network architectures and training pipelines for three-dimensional scene reconstruction from two-dimensional images. In other words, they solve the inverse problem of going from images of a scene (2D projections from different perspectives) to the scene these images are representing (the 3D object images were taken of). The method introduced in \cite{izzo_geodesy_2022} transfers this approach to the realm of geodesy, basically exchanging 2D images by gravitational forces and the 3D scene by the density field of a small body.
Thus, geodesyNets approximately solve the inverse problem of finding a gravitational body's mass density field given the gravitational force it exerts on other bodies at different locations, e.g., on a spacecraft along an orbit around it (\cref{fig:geodesy_arch}, left).

GeodesyNets are fully-connected neural networks with periodic activation functions (i.e. SIRENs, with $\omega_0 = 30$). The final layer uses the absolute value as an activation function to ensure densities are positive. The inputs to the network are Cartesian coordinates $x, y, z$, denoting a point within the three-dimensional space $V$ enclosing the body. In our case, the network consists of 4 fully-connected hidden layers with 300 neurons each. The network's output is the mass density of the small body, $\rho(x,y,z)$, at location $(x, y, z)$.
By using the neural network as a supplement for the actual mass density field of the small body, we calculate the expected gravitational force it enacts on other objects at a given location (e.g. a spacecraft along an orbit around it). The difference between the inferred gravitational acceleration and the ground truth value is evaluated in a custom loss function, which is minimized to improve the neural network's representation of the density field.
Both for reference models and the simulated memristor-based geodesyNets, we use 10,000 training epochs and a quadrature of 30,000. Though geodesyNets are generally applicable to any small, irregularly shaped body, training for this work is performed in all cases on the asteroid Eros. 

Different from G\&CNETs, which are pre-trained before deployment, geodesyNets continuously learn to fully characterize the body they are orbiting \cite{blazquez2023small} -- requiring many write operations while being deployed.
Thus, they would benefit significantly from a memristive implementation, which provides highly power-efficient write operations.

Alternative methods to perform the same or similar tasks on small bodies include spherical harmonics, using mascon models and polyhedral gravity models or a combination thereof \cite{werner_exterior_1997, wittick_mixed-model_2019, sebera_spheroidal_2016}, which geodesNetsy outperforms in terms of accuracy, needed prior assumptions and on-board applicability. As such we survey geodesyNets, as a representative method for this particular application.

\subsection{Memristor-based Neural Network Accelerators}\label{sub:memristor}
Originally introduced in 1971 \cite{chua1971memristor}-- but only first manufactured in 2011 \cite{williams2008we} -- memristors are an emerging technology that may enable neural networks to meet essential requirements in terms of accuracy, energy efficiency, radiation tolerance, and latency.
The memristor (a portmanteau of ``memory resistor'') is a two-terminal nonlinear resistive element, whose resistance $R$ (or conductance $G$, the inverse of the resistance) depends on the history of voltage across it. The method by which memristors switch their resistance (the so-called ``switching mechanism'') differs per technology used to implement the memristor, but is generally performed by one or more electrical pulses forming and rupturing some form of conductive filament, or changing a magnetic orientation \cite{zahoor_resistive_2020}.

\begin{figure}[htb]
    \centering
    \includegraphics[trim=0.0cm 0.5cm 0.5cm 0.0cm,width=1.0\columnwidth]{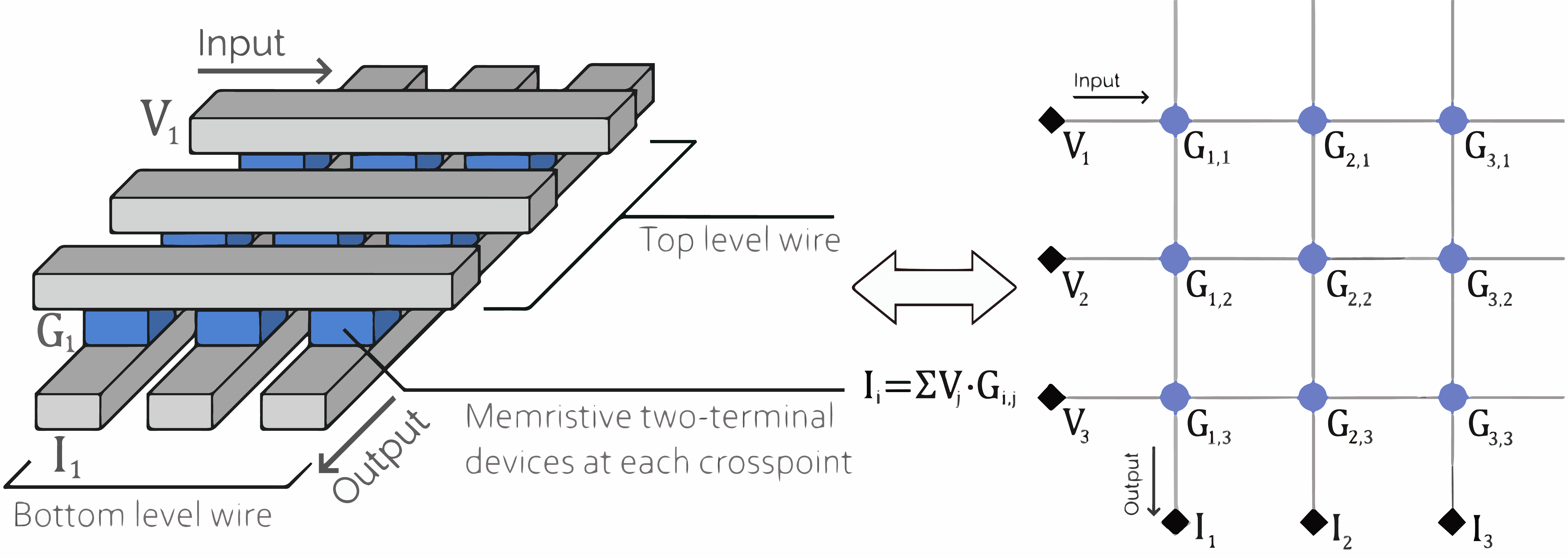}
    \caption{Memristors arranged in crossbar arrays, capable of performing matrix-vector-multiplication (MVM). Whereby a given output current $I_i$, is the result of a multiplication of input voltage $V_i$ and weight value--represented by conductance $G_{i,j}.$}
    \label{fig:mca}\vspace{-2.5mm}
\end{figure}

Memristor crossbars (shown in \cref{fig:mca}) can store multiple bits of information in each device, and are able to perform computations with very low energy and latency \cite{ankit_puma_2019}. If an input voltage is applied at the crossbar array's rows, we are able to get the result of a multiplication with the device's resistance as a consequence of Kirchhoff's law, thus performing matrix-vector-multiplication (MVM) in a crossbar array in one computational step. Each element of the crossbar (representing a weight, or element of the vector) is represented by two devices (2R, ``Two Resistors'') in a differential configuration (shown schematically in \cref{fig:differential_slicing}~(left). In this configuration, each neural network weight $w$ is encoded using a differential pair of memristive devices (commonly referred to as a 2R or "two-resistor" scheme), with associated conductances $G^+$ and $G^-$. The effective weight is defined as the difference between these conductances: $w \propto G^+ - G^-$.
In this configuration, one device encodes the positive component, and the other the negative component. Typically, either $G^+$ or $G^-$ is programmed to a conductance value within the supported range $[G_{\min}, G_{\max}]$, while the other is held at $G_{\min}$ (ideally 0).
These memristor crossbar arrays can be built using RRAM \cite{yu_rram_2021}, PCM \cite{wong_phase_2010} and other device technologies. As all calculations are performed in the analog domain, DACs and ADCs are required when interfacing with digital inputs our outputs. Although memristors may be implemented in a radiation resistant manner, it is important to note that the peripheral circuits, which include control blocks and the aforementioned DACs and ADCs, may need radiation hardening when a memristive accelerator is to be used in harsh environments \cite{hughart_comparison_2013, schumann_radiation_2022}. Through the experiments performed we wish to explore various configurations and parameters for a memristor-based neural network accelerator. For this we use a simulation setup which allows for modification of values such as the device properties and device technologies as necessary.

\begin{center}
    \begin{figure}[b!]
    \begin{subfigure}[t]{.5\linewidth}
      \centering
      \includegraphics[trim=0.22cm 0.22cm 0.1cm 0.22cm, width=0.695\columnwidth]{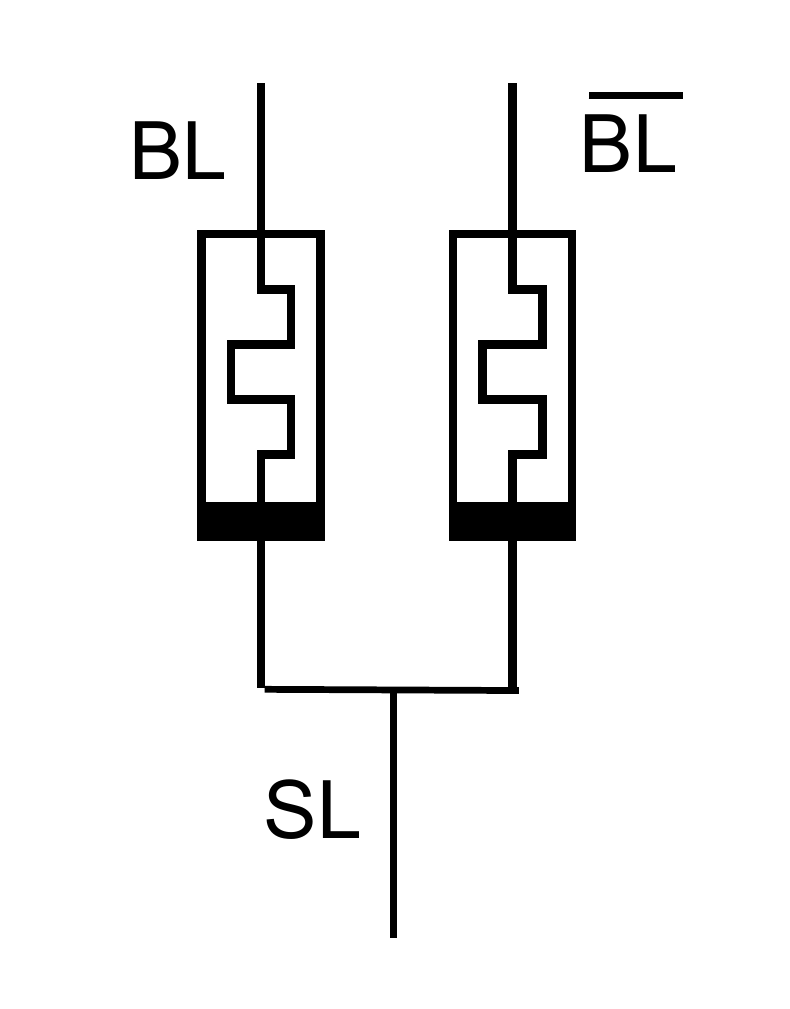}
    \end{subfigure}%
    \hfill
    \begin{subfigure}[t]{.50\linewidth}
      \centering
      \includegraphics[trim=0.22cm 0.22cm 0.1cm 0.22cm,width=1\columnwidth]{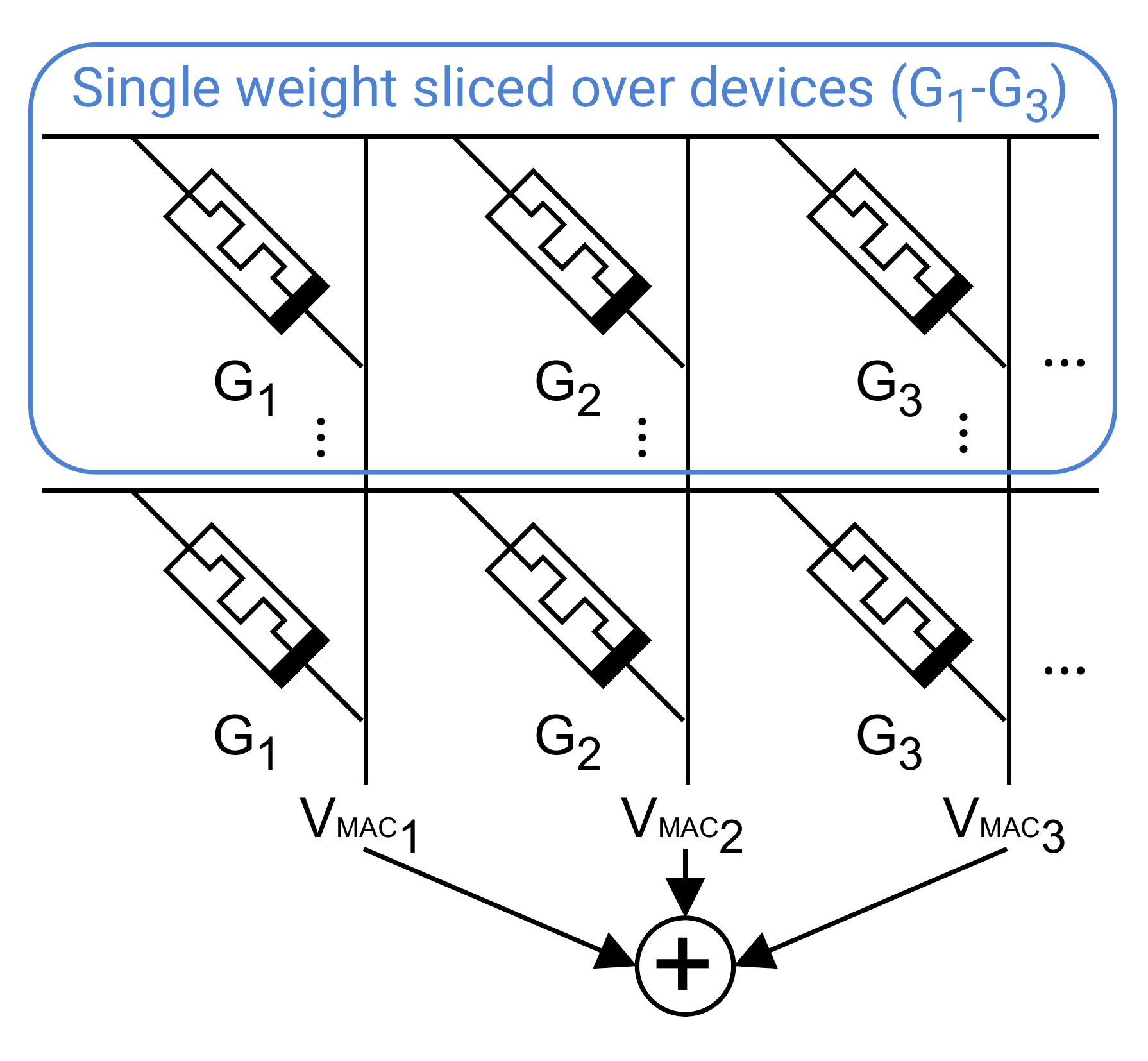}
    \end{subfigure}
    \caption{\textbf{(Left)} shows a schematic representation of 2R scheme commonly used for differential weight encoding, with BL (bit-line) and BLb often connecting to access transistors in a 2T2R (2 Transistor 2 Resistor) configuration. \textbf{(Right)} gives a diagram of bit slicing as being a technique wherein multiple multiple devices are used to represent one weight with higher precision.}
    \label{fig:differential_slicing}
    \end{figure}\vspace{-7.5mm}
\end{center}

\subsection{Simulation Setup}
The memristor-based neural network accelerator simulation setup used to perform the experiments is implemented using the \textit{IBM Analog Hardware Acceleration Kit} (IBM AI HW Kit for short) \cite{rasch_flexible_2021}, an open-source Python toolkit that is used to explore in-memory computing using analog devices in the context of neural networks  in a realistic manner. It is fully integrated within the PyTorch machine learning library for the simulation of training and inference of deep neural networks  built with PyTorch on analog crossbar arrays. It also features a variety of analog neural network modules such as fully connected layers and convolutional layers. Both analog training and hardware-aware training are supported. Furthermore, it is highly customizable in every aspect, ranging from mapping algorithms to supported analog layer types. As part of our contribution to the toolkit, bit-slicing \cite{gallo_precision_2022} (as linear bit sliced layers) has been implemented with the ability to vary the number of slices arbitrarily. 
Bit-slicing is the use of multiple devices to represent the value of one weight in the neural network, with each device representing a ``slice'' of a complete weight. \Cref{fig:differential_slicing}~(right) shows how bit-slicing would be implemented in practice on a circuit level.

In order to achieve realistic simulation of inference accuracy degradation, the IBM AI HW Kit also provides a specific inference configuration that adds carefully calibrated conductance-dependent programming noise, weight read noise and conductance drift based on a 1M PCM device array. 
The PCM devices exhibit approximately 2\% read noise on its programmed conductance on average. The same has been done for RRAM devices, based on data from Wan et al.'s work on RRAM-based in-memory computing chips \cite{wan_compute--memory_2022}. The RRAM exhibits 1\% read noise on its programmed conductance on average.
Conductance drift is a non-ideal behavior of memristors, meaning they may lose their programmed value over time even without any applied voltage. ``Refreshing'' the devices to restore the weights is possible, but consumes energy \cite{baroni_tackling_2021}.

The architecture of the hardware computing the forward pass of the network is as follows: the network consists of multiple layers, each consisting of one or multiple tiles, which contain the memristor crossbar arrays and associated periphery. The weights are mapped to the devices in each tile in a differential configuration, with one device representing a positive synaptic weight and another a negative component, with one of either being set to 0. The peripheral circuitry consists of digital-to-analog converters (DACs, set to a 7-bit resolution) on the input and analog-to-digital converters (ADCs, set to a 9-bit resolution) on the output. It also simulates IR-drop (voltage dropping as current passes through a resistor) and noise from the peripheral circuits (such as from amplifiers used in the ADCs). The result of the computed MVM is then passed to the activation function, which then passes its output to the next layer. 

\subsection{Device fault and non-ideality mitigation}
Within this simulation setup, we propose the concept of ``temporal averaging'' to deal with hardware-related stochasticity. 
The core idea is as follows: given a single layer in the neural network, its output is calculated multiple times and averaged before being sent to the next layer. 
This way, the output of the layer resembles more closely the intended, noise-less and variability free behavior.
Moreover, amplification of noise due to propagation through the network is directly inhibited.
For instance, in case of noise of a Gaussian distribution on the weights, the standard deviation of the mean output will decrease by a factor $\frac{1}{\sqrt{N}}$, where $N$ is the number of repeats we are performing.
Assuming that the means of the weight distributions are the values we intended to program, averaging produces layerwise outputs that are closer to the intended ones without noise.

A large benefit of this mitigation technique is the simplicity of potential implementations in hardware. Combining a shift-register and an adder to capture and combine the result of multiple runs of the layer (or network), which can then be averaged by simply shifting the captured result right by the amount of results captured (thus limiting the system to repeats of powers of 2).
It further utilizes the fact that forward passes are cheap both in terms of energy and latency on memristive devices -- although excessive averaging will add noticeable costs that might detract from the benefits of memristors.

As an alternative to layerwise averaging, we explore commonly used bit slicing, with the final weight being the average of the slices.
Different from temporal averaging, where averaging was performed for the same device, here we spatially average over different devices, albeit with related read/write behavior.
Thus, we expect a similar reduction in noise levels using this technique.

Lastly, to deal with hardware-related non-idealities (e.g. device faults), hardware-aware (HWA) training is used to train the memristor-based neural networks. In HWA training, weight updates are calculated in digitally (i.e., not on-chip) based on inference run on hardware \cite{esser2016,schmitt2017neuromorphic,kungl2019accelerated,rasch_hardware-aware_2023}. In our work, an ideal backward pass is assumed, with the non-idealities only affecting the forward pass. Weight updates derived from this process are then applied to the simulated memristive neural network, which is repeated for all training epochs. In a produced chip, the hardware-aware training could be substituted by a digital section of the chip, rather than on a separate machine. The code and data to reproduce the results in this paper is available on Github \cite{github_repo}.

\vspace{-2.5mm}
\section{Results}\label{sec:results}
We perform a thorough set of experiments using the simulation setup outlined in \cref{sub:memristor} to assess the current suitability of memristive neural network accelerators implemented with the device technologies introduced in \cref{sec:intro}.
Moreover, we demonstrate the feasibility of the proposed noise mitigation techniques, and study the effect of device imperfections (faulty devices and conductance drift) on task performance.
In particular, we study:
\begin{enumerate}
    \itemsep0em 
    \item The impact of linear bit-slicing on performance, from 1 slice to 16 slices.
    \item The impact of temporal layerwise averaging, with averages (repeats) ranging from no repeats (1) to 64, in intervals matching powers of 2.
    \item The performance under various levels of device degradation (as a ratio of devices stuck at any value between $G_{min}$ and $G_{max}$, randomly distributed), in a realistic range from 1\% to 10\% \cite{kim_memristor_2024}.
    \item Performance recovery through re-training the neural networks post-degradation, i.e., using device-aware training.
    \item The effect of conductance drift over time on performance (from $t=0s$, to $t=48\cdot3600s$).
\end{enumerate}
For both the navigation and geodesy task, the performance is defined as the average loss of the test set.
All experiments are performed for both the PCM and RRAM devices. 
Aside from the experiments concerning the number of devices per weight (slices), all experiments are performed with 1 slice per weight for the G\&CNETs, and 4 slices per weight for geodesyNets. These values were chosen on the basis of the effectiveness of the slices during the experiments concerning the linear bit-slicing. 

\subsection{Memristor-based G\&CNETs}


\begin{center}
    \begin{figure}[b!]
    \begin{subfigure}[t]{.5\linewidth}
      \centering
      \includegraphics[trim=0.22cm 0.22cm 0.1cm 0.22cm, width=0.995\columnwidth]{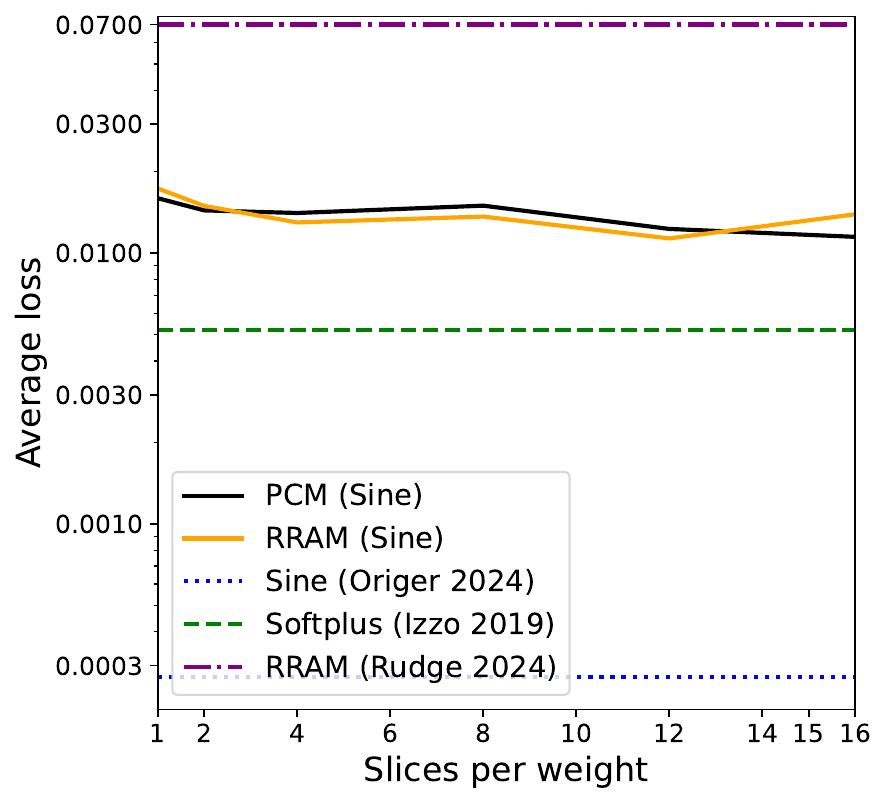}
    \end{subfigure}%
    \hfill
    \begin{subfigure}[t]{.50\linewidth}
      \centering
      \includegraphics[trim=0.22cm 0.22cm 0.1cm 0.22cm,width=0.995\columnwidth]{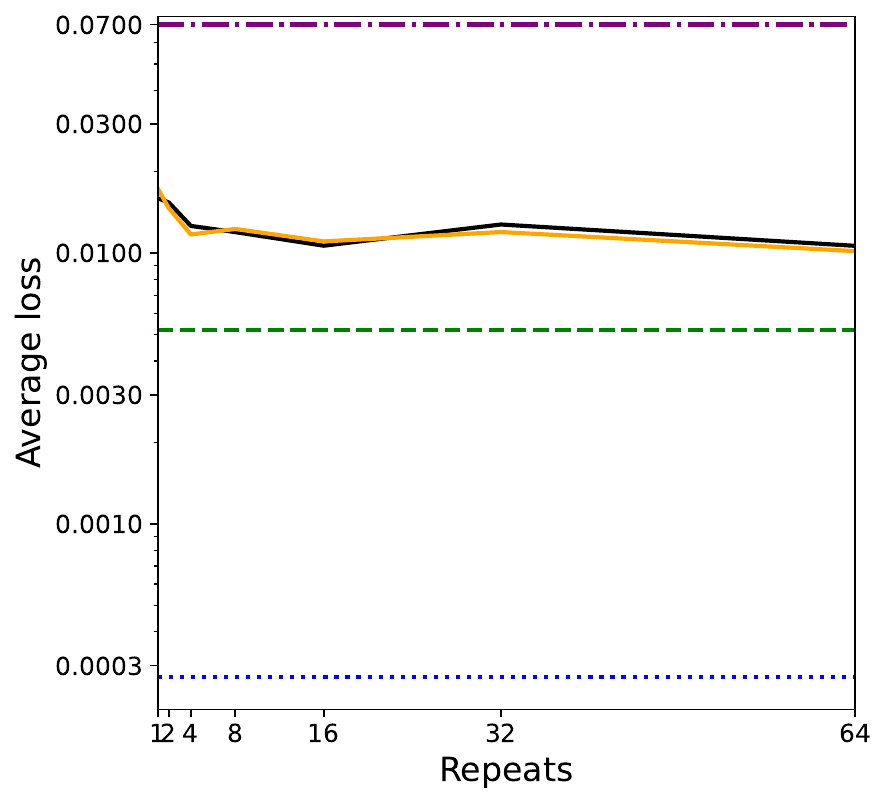}
    \end{subfigure}
    \caption{Average loss over the test dataset of model predictions plotted against \textbf{(left)} slices (with repeats set to 1) and \textbf{(right)} number of temporal averages (with slices set to 1). For comparison, we also show the best result obtained with simulated RRAM for neural networks with softplus activation function (dash-dotted), as well as the digital baseline for softplus (dashed) and sine activation function (dotted).}
    \label{fig:gecnet_slicesrepeats}
    \end{figure}\vspace{-7.5mm}
\end{center}

In \cref{fig:gecnet_slicesrepeats}~(left), we see that the number of slices has only a minor effect on the accuracy of the network, which is why we default to only $1$ slice per weight for the G\&CNET. 
The same is true for temporal averaging (\cref{fig:gecnet_slicesrepeats}, right).
However, by switching from softplus activation functions to sine functions, we see an improvement from $0.07$ to almost $0.01$ -- although the digital baseline also improves by almost one order of magnitude.
Nevertheless, the memristor-based G\&CNET gets close to the performance level of the digital baseline using softplus activation functions.


\begin{center}
    \begin{figure}[ht]
    \begin{subfigure}[ht]{.5\linewidth}
      \centering
      \includegraphics[trim=0.22cm 0.22cm 0.1cm 0.22cm, width=0.995\columnwidth]{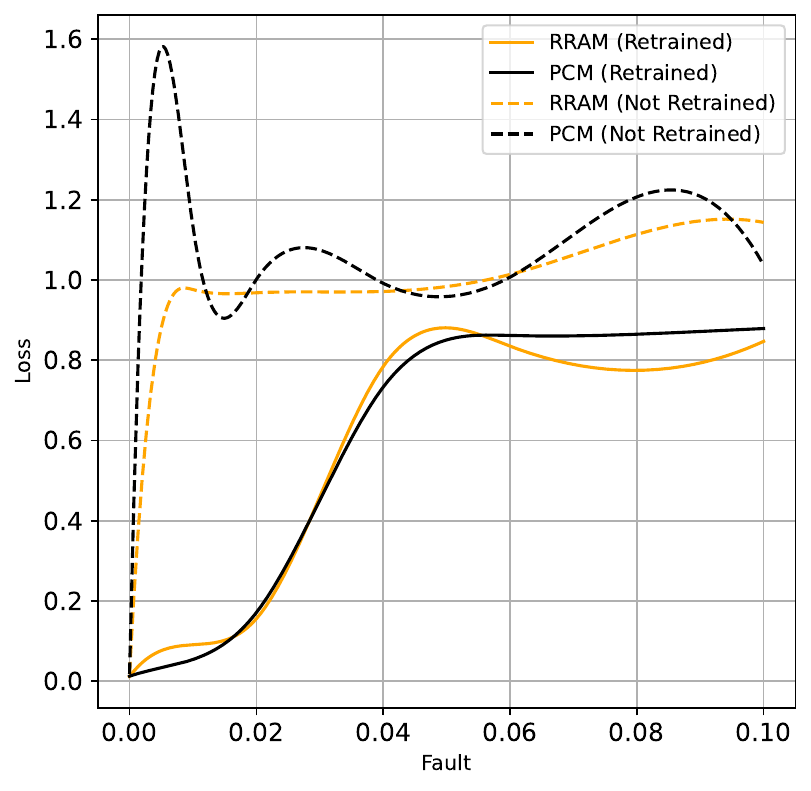}
    \end{subfigure}%
    \hfill
    \begin{subfigure}[ht]{.50\linewidth}
      \centering
      \includegraphics[trim=0.22cm 0.22cm 0.1cm 0.22cm,width=0.995\columnwidth]{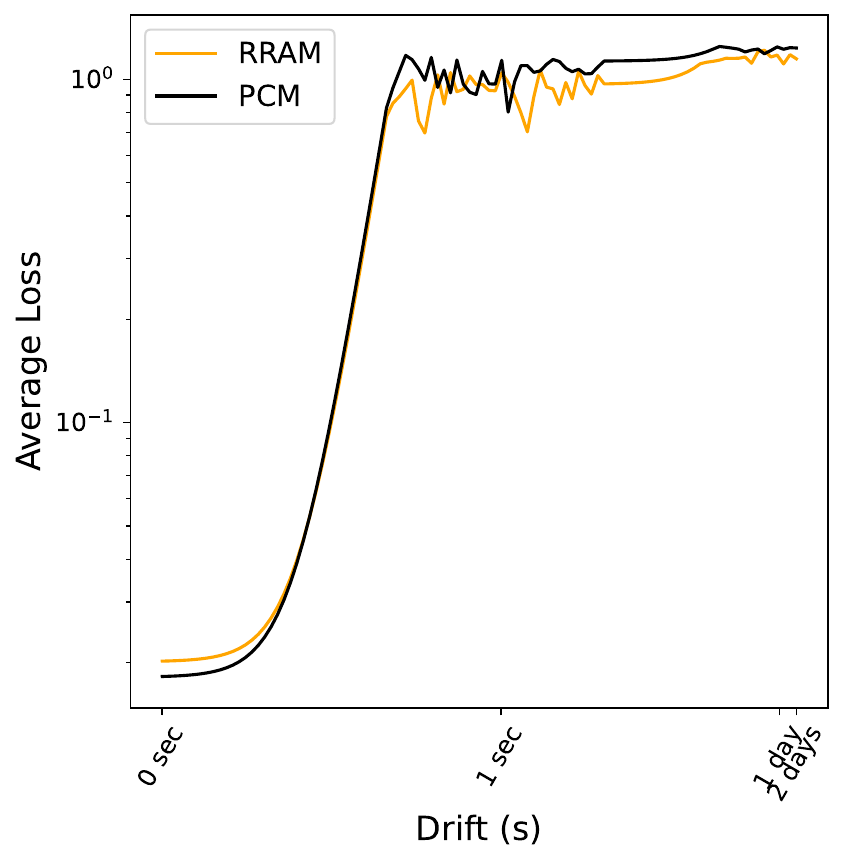}
    \end{subfigure}
    \caption{\textbf{(left)} Loss for increasing fault ratios. The two sets of lines depict the PCM and RRAM neural networks  as affected by faulty devices with (line) and without (dashed) retraining. \textbf{(right)} The effect of conductance drift on the performance of the network.}
    \label{fig:gecnet_faultsdrift}
    \end{figure}\vspace{-7.5mm}
\end{center}

\Cref{fig:gecnet_faultsdrift} shows a rapid decline in network performance as device failures increase, with fault ratios above 2–3\% already presenting significant challenges for re-training. In the current model, even a small number of faults severely impacts performance, with the non-retrained network exhibiting substantial degradation. Compared to the softplus-based architecture from \cite{rudge_guidance_2024}, this suggests that the G\&CNET architecture with periodic activation functions is highly vulnerable to disturbances. For example, when comparing 10\% device faults without re-training in \cref{fig:gecnet_faultsdrift} and with re-training, we see that the network recovers significantly from a loss of around $1.1435$ to $0.87$ -- but in the softplus case this was a recovery from around $0.34$ to $0.086$.

\cref{fig:gecnet_faultsdrift}~(right) shows the effect of conductance drift on the loss of the network, which is similar to the effect observed for device faults. Before $t=0.5s$ the network is mostly unaffected, after which performance deteriorates quickly. In previous network architectures, the softplus version of G\&CNETs \cite{rudge_guidance_2024} was able to hold performance much longer, and as will be shown in the next section, our results for GeodesyNet are more in line with those  results (see \cref{sec:geodesy_results}).

\begin{figure}[t]
    \centering
    \includegraphics[width=0.75\linewidth]{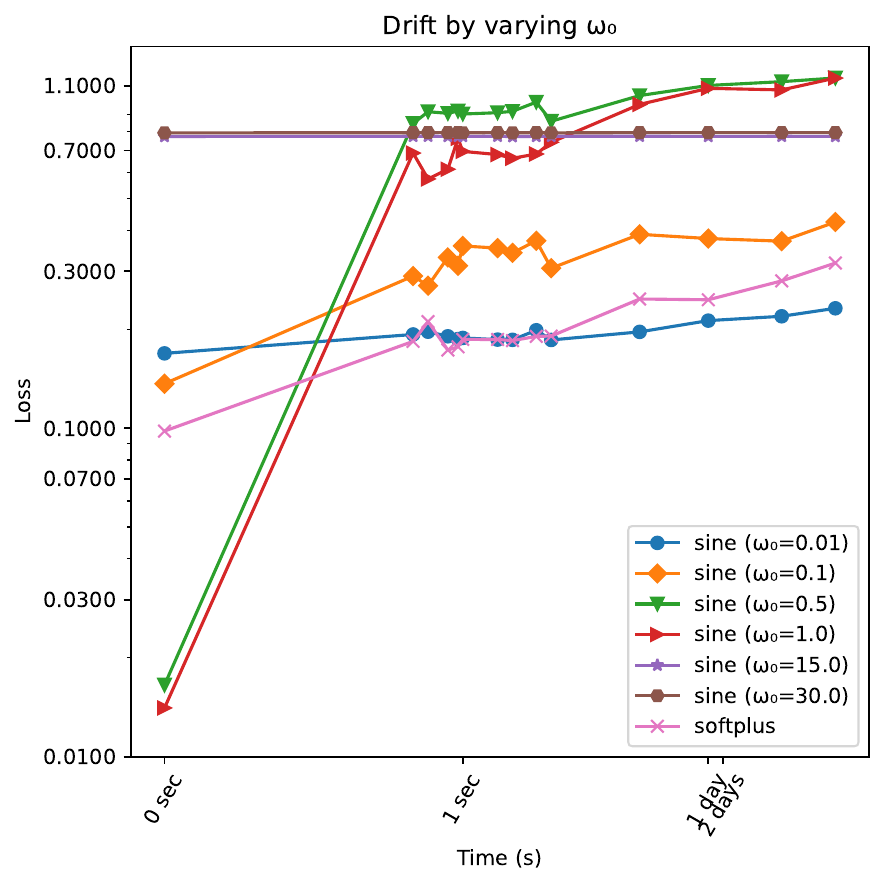}
    \caption{A plot showing the performance under drift at various values of $\omega_0$. Showing that at a high enough $\omega_0$, the system does not learn at all, and at a small enough $\omega_0$ though robust to drift, the initial performance is very close to the original softplus implementation.}
    \label{fig:drift_omega}
\end{figure}

We observed that the G\&CNETs with sinusoidal activations exhibit a significantly stronger degradation under conductance drift and device failure compared to their softplus-based counterparts. To investigate this behavior, we analyzed the sensitivity of the networks with respect to perturbations in the weights. In an effort to create a more detailed analysis, we also trained the network with a variety of $\omega_0$ values, ranging from $\omega_0=30$ to $\omega_0=0.01$ and determined the Lipschitz constant of each of these networks, and also that of the softplus-based network and GeodesyNet. The results of these tests are given in \cref{fig:drift_omega}. The Lipschitz constant was estimated numerically by two methods: (1) computing the average gradient norms using automatic differentiation over the dataset, and (2) evaluating $|f(x)-f(x')|/||x-x'||$ for a large number ($>1500$) of random input pairs. We observe that the Lipschitz constant seems correlated to the sensitivity of a given network to drift. At $\omega_0=30$, this value is 1548, at $\omega_0=1.0$ it is 355, and finally at $\omega_0=0.1$ and $\omega_0=0.01$ it results in 27 and 3 respectively (all have been rounded). With $\omega_0=0.01$ showing poorer performance at $t_0$ than $\omega_0=1.0$, but much stronger robustness against drift. This phenomenon is discussed further in Section~\ref{sec:discussion}.

Finally, the spacecraft control obtained from the current best analog network (1 slice, RRAM, no faults nor drift, 300 epochs at 64 repeats for temporal averaging) for a single interplanetary transfer is shown in \cref{fig:phitheta}, together with the control produced by the digital baseline. The loss of this network is approximately $0.01018$.  More specifically, we show the target spherical coordinates ($\theta$ and $\phi$) as predicted by both the digital and analog neural networks  -- illustrating that the analog model is capable of performing transfers after training, although with noisier control than the digital model.
%

\begin{center}
    \begin{figure}[t]
    \begin{subfigure}[ht]{.5\linewidth}
      \centering
      \includegraphics[clip, trim=0.1cm 0.26cm 7.2cm .26cm, width=1\columnwidth]{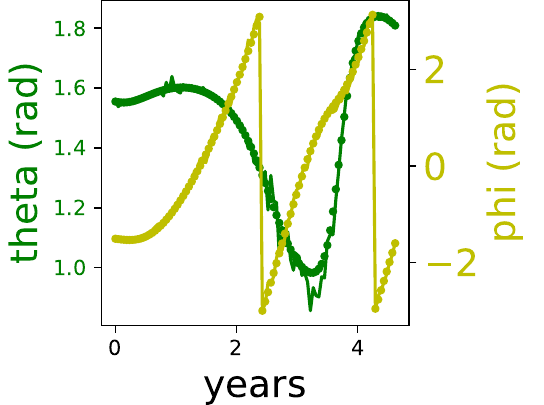}
    \end{subfigure}%
    \hfill
    \begin{subfigure}[ht]{.5\linewidth}
      \centering
      \includegraphics[clip, trim=9.6cm 0.26cm 7cm .26cm,width=0.93\columnwidth]{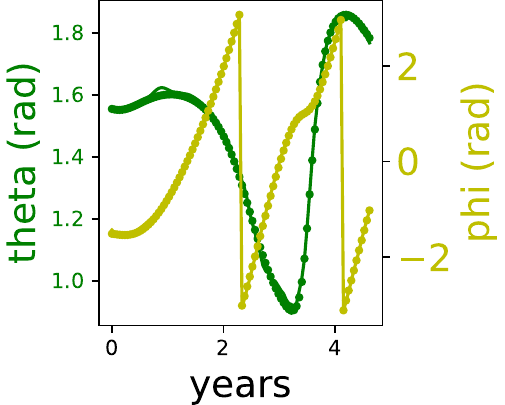}
    \end{subfigure}
    \caption{The network prediction transformed into spherical coordinates $\theta$ and $\phi$ (lines) compared to the optimal ground truth (points). Both the analog (RRAM, left) and digital (right) models are shown.}
    \label{fig:phitheta}
    \end{figure}\vspace{-7.5mm}
\end{center}

\subsection{Memristor-based geodesyNets}\label{sec:geodesy_results}

\begin{center}
    \begin{figure}[ht!]
    \begin{subfigure}[ht!]{.5\linewidth}
      \centering
      \includegraphics[trim=0.22cm 0.22cm 0.1cm 0.22cm, width=0.995\columnwidth]{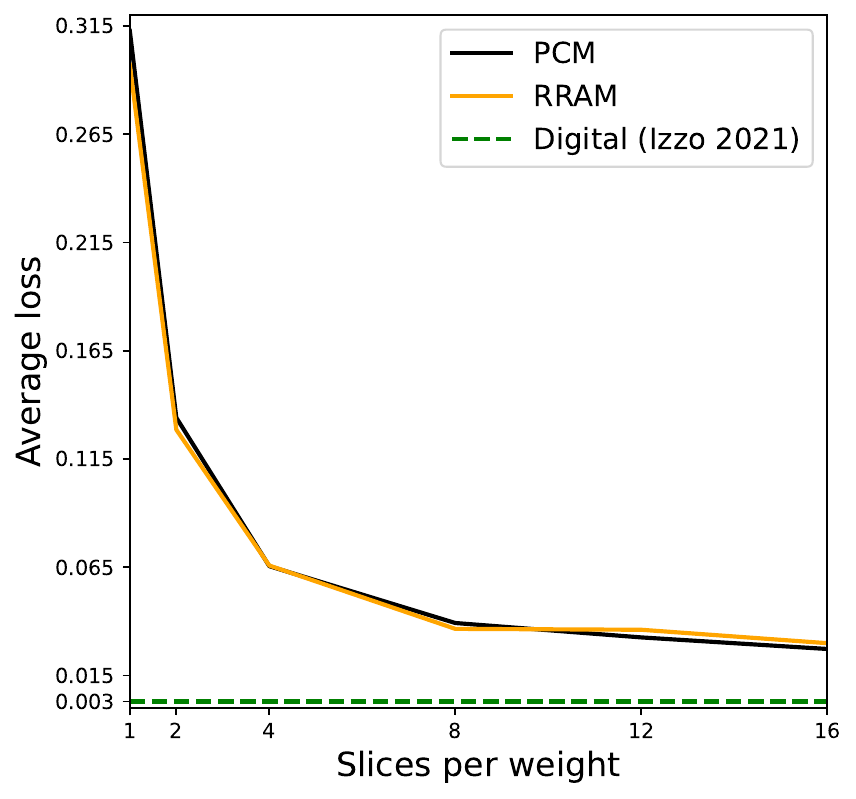}
    \end{subfigure}%
    \hfill
    \begin{subfigure}[ht!]{.50\linewidth}
      \centering
      \includegraphics[trim=0.22cm 0.22cm 0.1cm 0.22cm,width=0.995\columnwidth]{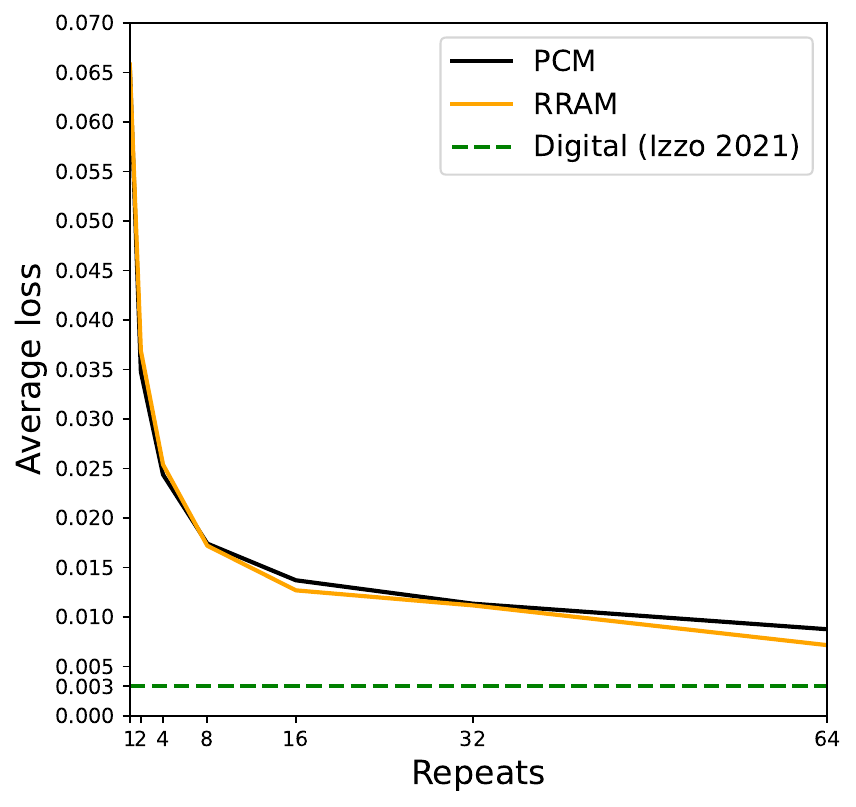}
    \end{subfigure}
    \caption{\textbf{(left)} Loss of model predictions plotted against slices. Digital baseline is shown as a dashed line. \textbf{(right)} The effect of using temporal averaging on the network. Each experiment was run with 4 slices per weight. Note that the y-axes have different ranges.}
    \label{fig:geodesy_slicesrepeats}
    \end{figure}\vspace{-7.5mm}
\end{center}

Different to G\&CNets, without bit-slicing or temporal averaging the network completely fails to solve the inverse problem during learning.
This is not only seen in the large loss of around $0.36$ (compared to the digital baseline of $0.003$), but also reflected in the reconstructed density field which fails to capture any structure of the asteroid (\cref{fig:geodesy_worstokay}, left).

In \cref{fig:geodesy_slicesrepeats} we show that increasing the number of slices drastically improves network performance, with diminishing returns for larger number of slices. Similarly, using temporal averaging (repeats) strongly enhances performance (\cref{fig:geodesy_slicesrepeats}, right). Of note is that at 4 slices, the loss has reached approximately a value of $0.0655$, whereas at 4 repeats (and 4 slices) the best loss reaches $0.0253$.
This highlights that both techniques can be used complementarily to further enhance performance of the final model.
In total, we reach a performance of $0.008$ for PCM and $0.007$ for RRAM (4 slices, 64 repeats), getting to the same order of magnitude in loss as the digital baseline ($0.003$).

\begin{center}
    \begin{figure}[ht]
    \begin{subfigure}[t]{.5\linewidth}
      \centering
      \includegraphics[trim=0.22cm 0.22cm 0.1cm 0.22cm, width=0.995\columnwidth]{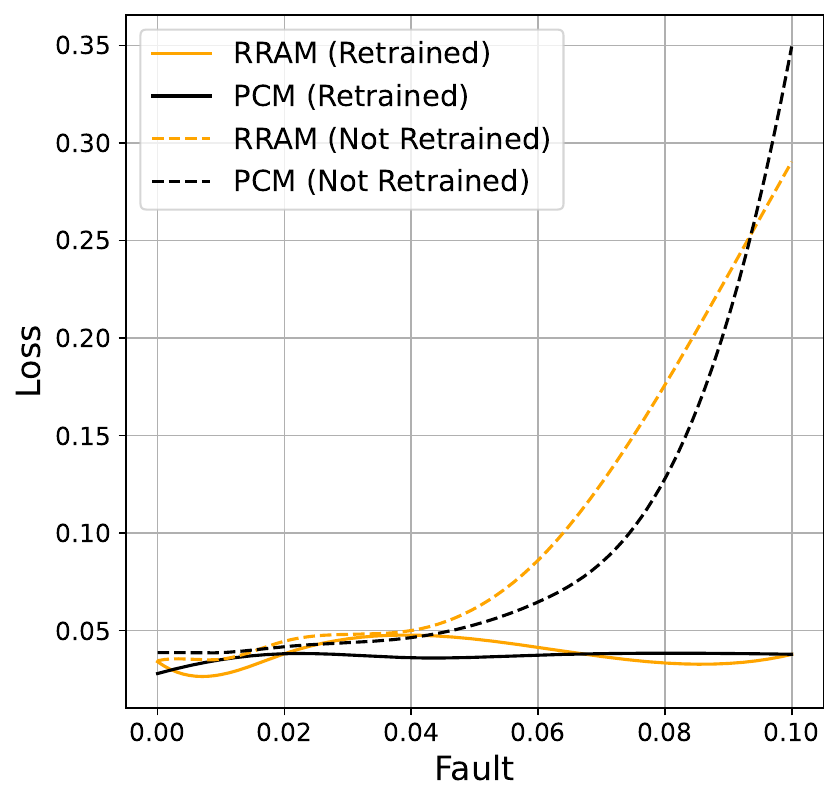}
    \end{subfigure}%
    \hfill
    \begin{subfigure}[t]{.50\linewidth}
      \centering
      \includegraphics[trim=0.22cm 0.22cm 0.1cm 0.22cm,width=0.995\columnwidth]{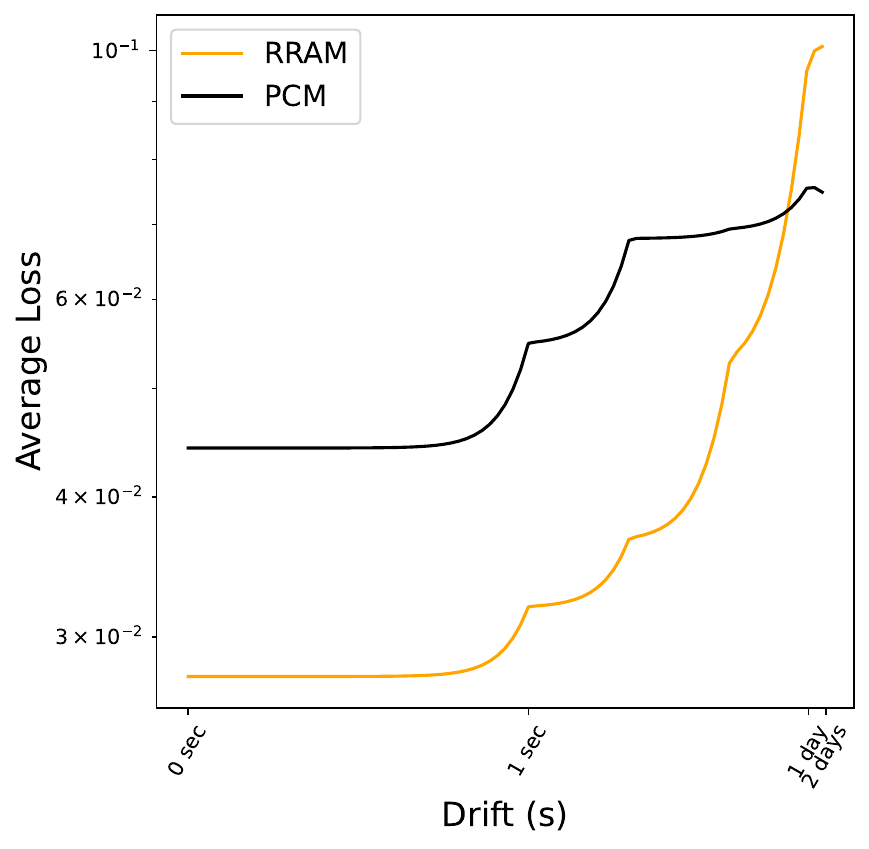}
    \end{subfigure}
    \caption{\textbf{(left)} Average loss of the model while varying the number of faulty devices in the neural network, up to 10\%. The effect of post-degradation re-training is also shown. \textbf{(right)} The system's performance under drift, up to 2 days.}
    \label{fig:geodesy_driftfaults}
    \end{figure}\vspace{-7.5mm}
\end{center}

\cref{fig:geodesy_driftfaults}~(left) shows the performance of the neural network under the effects of device degradation. The network is fairly robust up to 4-6\% device degradation, even without compensatory re-training. After this point, re-training is able to maintain performance within the same range of loss, up to 10\% (from the original performance of around $0.0279$ to $0.0377$ after re-training, with the uncompensated loss at $0.321$).  In \cref{fig:geodesy_driftfaults}~(right) we see how the network deals with conductance drift. At approximately 1 second, performance is fairly stable, with performance degrading rapidly as we approach 1 day. This loss development as the conductance drift worsens is more in line with expectations, as opposed to the severe impact seen in the G\&CNET case, where even the slightest drift or device degradation completely degraded performance. We see also the difference in the onset of the largest effects of the drift between PCM and RRAM, with RRAM's slope being more gentle, but not tapering off until we reach 2 days of drift.


\begin{center}
    \begin{figure}[t!]
    \begin{subfigure}[h!]{.5\linewidth}
      \centering
      \includegraphics[trim=0.5cm 0.5cm 0.5cm 1cm, clip, width=0.995\columnwidth]{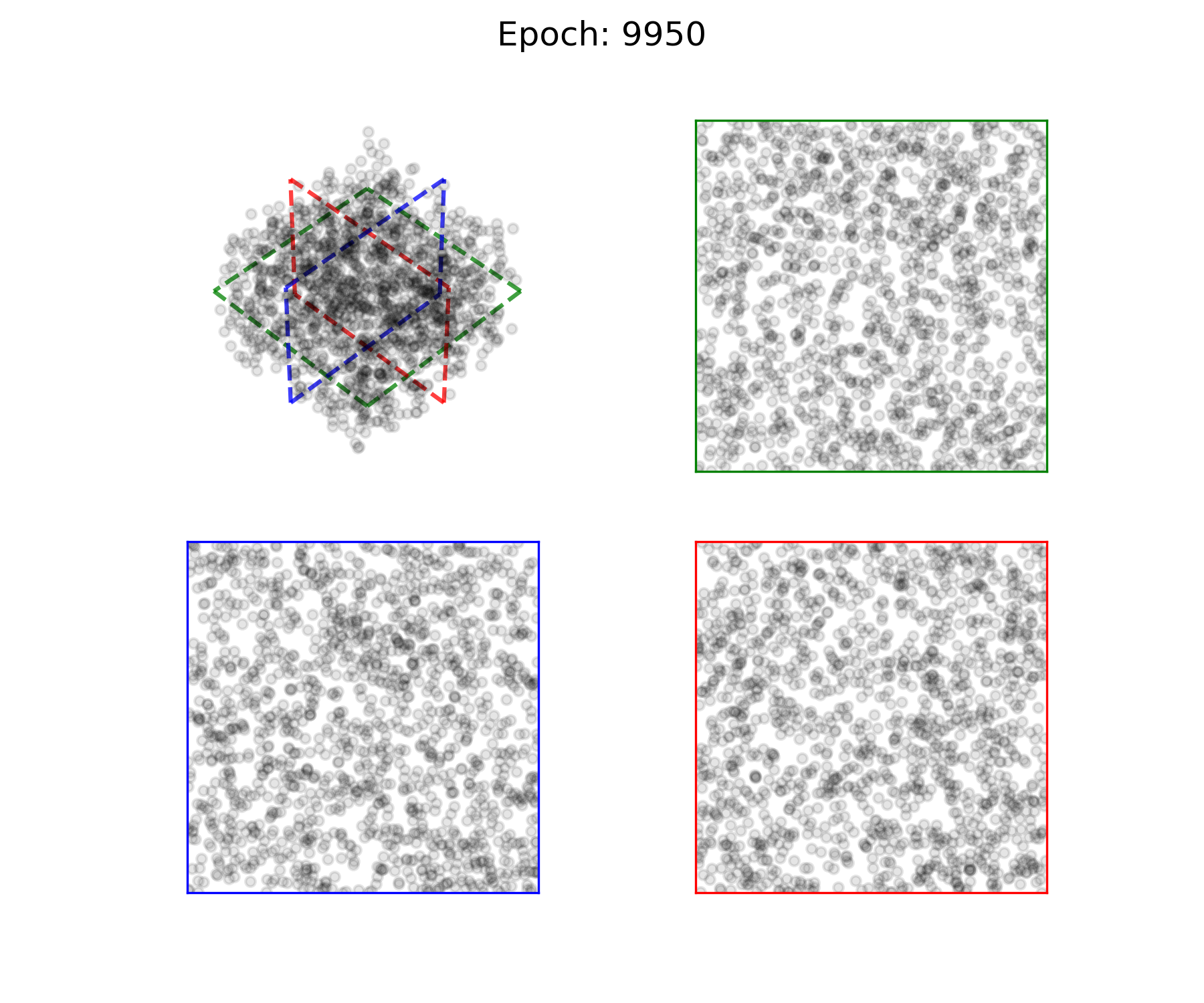}
    \end{subfigure}%
    \hfill
    \begin{subfigure}[h!]{.50\linewidth}
      \centering
      \includegraphics[trim=0.5cm 0.5cm 0.5cm 1cm, clip,width=0.995\columnwidth]{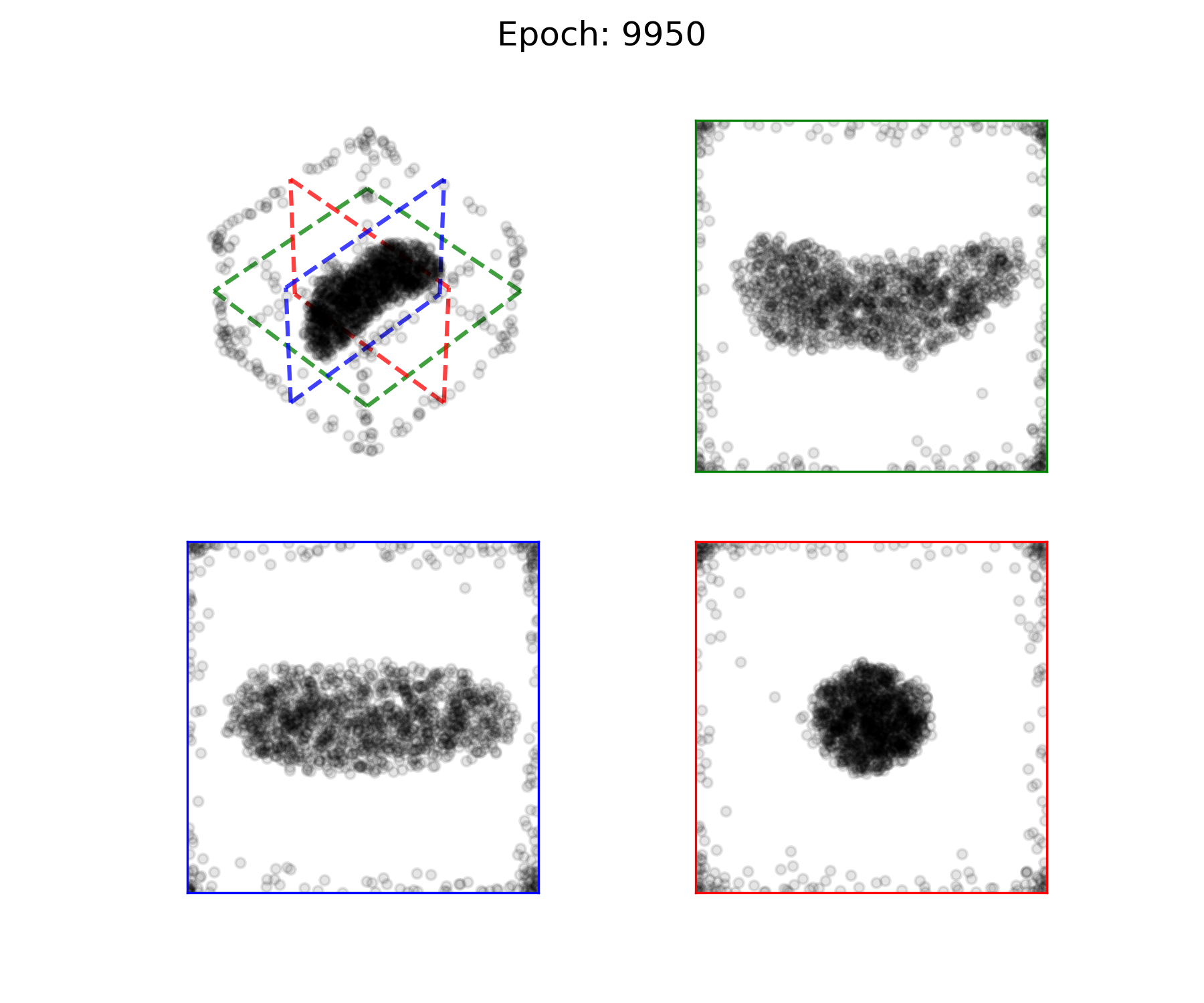}
    \end{subfigure}
    \caption{\textbf{(left)} The predicted asteroid density in a model featuring no temporal averaging and only 1 slice, showing its inability to learn at this level of noise and variability. \textbf{(right)} With a moderate level of noise mitigation (4 slices and 4 repeats for the temporal averaging), the model learns the density field, but struggles with cleaning up the edges.}
    \label{fig:geodesy_worstokay}
    \end{figure}\vspace{-7.5mm}
\end{center}

\begin{figure}[t]
    \centering
    \includegraphics[trim=0.5cm 0.5cm 0.5cm 1cm, clip, width=0.9\linewidth]{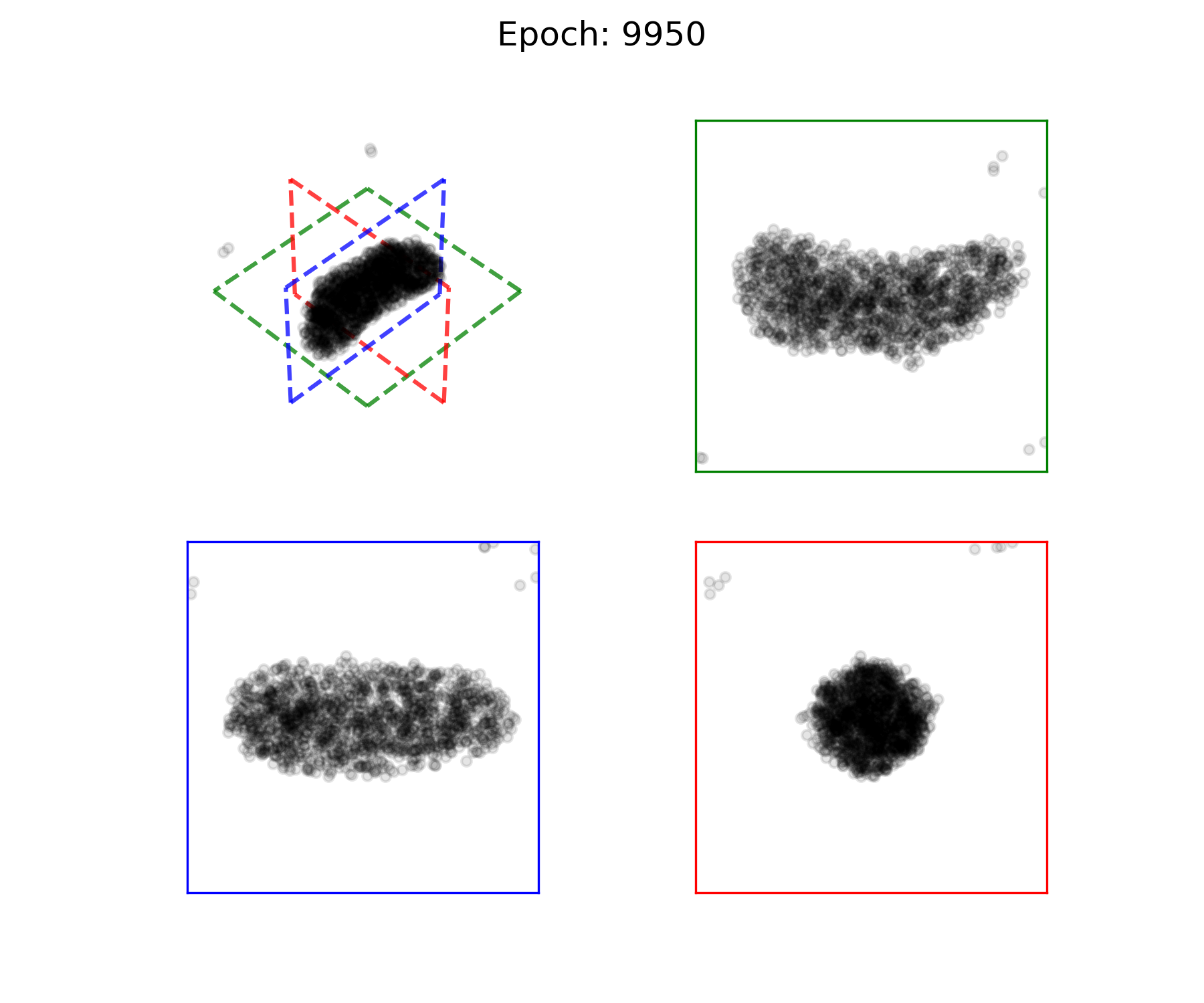}
    \caption{The plot of the best-performing model with 4 slices and 64 repeats. It reaches a final loss of approximately $0.008$. }
    \label{fig:geodesy_best}
\end{figure}

\Cref{fig:geodesy_worstokay} compares two neural network models trained to learn the shape and density distribution of the asteroid Eros. The left model, which does not include the mitigation techniques outlined in \cref{sec:methods}, does not learn at all, finishing training with a high loss of 
$0.36$. 
The learned density field is significantly improved (with a loss of $0.024$) when using 4 slices and 4 repeats, as shown in \cref{fig:geodesy_worstokay}~(right). 
Finally, in \cref{fig:geodesy_best} we show the best performing model, with 64 repeats for temporal averaging and 4 slices per weight. It converges to a final loss of approximately $0.008$ and clearly shows the shape and density of Eros, with errant densities at the edges cleaned up.
This demonstrates that with temporal averaging and slices, the memristor-based neural network is capable of accurately learning Eros’ structure from scratch in under 10,000 epochs.

\section{Discussion}\label{sec:discussion}

Using simulations based on the IBM AI HW KIT, we have demonstrated that memristor-based neural networks are capable of reaching promising performance levels on state-of-the-art onboard AI tasks. In some respects, our simulations may be overly pessimistic, as advances in technology -- such as improved RRAM devices \cite{falcone_all--one_2025} --  will produce better results than the devices we used in this study (mostly limited by which devices were modeled in the IBM AI HW KIT during the time of this study). In other respects, they may be too optimistic, as a reality gap always remains -- especially when modeling microelectronics on an abstract level.
This is particularly true regarding the extent to which non-idealities of devices are captured and modeled. An example limitation of the simulation is that the effect of non-ideal transistor matching is not included. 

The simulated memristor-based neural networks reach performance levels close to those of their digital counterparts, although a significant gap (around a factor of 2) remains.
This performance gap appears to be primarily driven by variability during inference and training, as evidenced by the effectiveness of bit-slicing and temporal averaging. Not only is temporal averaging effective by itself and when combined with bit-slicing, its benefits also extend to the simplicity of its potential implementation in hardware.
Since through averaging, variability is only reduced by a factor of $1/\sqrt{N}$ (for $N$ repeats), a future venue is to look into more efficient mitigation techniques compatible with memristor crossbar arrays, e.g., error correcting codes.

Both drift and device degradation have a significant impact on the performance of G\&CNETs, with
the performance degrading much more steeply than in the case of the softplus-based neural network architecture presented in \cite{rudge_guidance_2024}, despite the initial significantly lower loss. GeodesyNets, which use a similar SIREN architecture, do not suffer from issues with drift and device degradation, possessing sufficient robustness to these non-idealities.

This warranted further research and experimentation, particularly regarding the choice of $\omega_0$. The selected $\omega_0=1.0$ for G\&CNETs deviates from the customary $30.0$. As such, we investigated the effect of different $\omega_0$ values on the network's response to drift (as outlined in \cref{sec:results}). This behavior appears linked to the network’s Lipschitz constant: the more reactive a network’s output is to its input, the more it also amplifies perturbations due to weight drift or noise.
We find that G\&CNETs with higher $\omega_0$ values exhibit higher Lipschitz constants and are more sensitive to drift, whereas networks with lower $\omega_0$ are more robust but also less expressive. Interestingly, although GeodesyNets also use sinusoidal activations, they exhibit stronger robustness. This could be because the learned density field changes more gradually in space, and the integral over many points averages out local steepness. Furthermore, the Lipschitz constant of the GeodesyNets are in line with those of the softplus-based G\&CNETs, which exhibit better robustness to drift noise.
Normalization techniques may help address this, but further research is required to confirm this.
These findings highlight a crucial consideration in designing networks for memristive applications: loss alone is not a sufficient performance metric. A more holistic approach is required, as networks may train well but still exhibit poor robustness under important conditions such as drift and device degradation.
This underscores not only the need for thorough post-training validation, but training procedures and network architectures that exhibit robust parameters.

There are many avenues to expand this work.
One direction of interest is adding energy estimation within the simulation setup to compare the energy-consumption of memristor-based networks with conventional hardware. Further studies on network architectures and hyperparameters are also needed, particularly to explore the trade-offs between temporal averaging and slicing in terms of latency, energy, area, and accuracy in real hardware implementations. For geodesyNets, the current 10,000-epoch training run, chosen due to the prolonged run-times when simulating memristive devices, is relatively short compared to the original 25,000-epoch training, affecting accuracy. Additionally, further experimentation is needed to assess the impact of $\omega_0$ in periodic activation functions, especially in the case of G\&CNETs.

To extend the sandbox of models that can inform the design and development of memristive accelerators for space, additional space applications should be explored. With geodesyNets successfully implemented, additional applications using neural implicit representations are suitable candidates for memristors, such as Neural Radiance Methods for Lunar Terrain Modeling \cite{kints_neural_nodate}. Adaptive compression algorithms \cite{asiyabi_complex-valued_2023} are another candidate, optimizing SAR imagery compression and enabling in-field continuous learning. In such cases, memristors' ability to perform low-power inference and in-field training can prove particularly suitable. Another relevant point of interest is also the impact of on-chip training (as opposed to hardware-aware training) on geodesyNets and other networks that learn in the field, as on-chip training may affect the network's ability to learn properly.

Finally, this study is entirely simulation-based and thus inherently constrained by the data, models, and assumptions used. While great effort has been made to ensure realistic conditions, the simulation-to-reality gap remains. In future work we intend to address this by designing, fabricating, and characterizing hardware to assess the feasibility of memristor-based neural network accelerators for on-board AI.

To conclude, coupled with the latency, power, and energy reductions enabled by memristors \cite{bonnet_bringing_2023} as well as their radiation hardness \cite{amrouch_towards_2021}, our results provide a promising outlook for enabling efficient and reliable on-board AI using memristors. 

\section*{CRediT authorship contribution statement}
\textbf{Zacharia A. Rudge}: Conceptualization, Methodology, Software, Validation, Formal analysis, Investigation, Data curation, Writing – original draft, Writing – review \& editing, Visualization.
\textbf{Dominik Dold:} Conceptualization, Methodology, Software, Investigation, Visualization, Writing – review \& editing, Supervision.
\textbf{Moritz Fieback:} Conceptualization, Supervision.
\textbf{Dario Izzo: }Conceptualization, Methodology, Supervision, Writing – review \& editing.
\textbf{Said Hamdioui:} Conceptualization, Supervision.

\section*{Declaration of competing interest}
The authors declare that they have no known competing financial interests or personal relationships that could have appeared toinfluence the work reported in this paper.

\section*{Acknowledgments}
Z.R. acknowledges funding from the European Space Agency's Open Space Innovation Platform (contract number 4000140774).
D.D. was supported by the Horizon Europe's Marie Skłodowska-Curie Actions (MSCA) Project 101103062 (BASE).
We also acknowledge the use of the open source software IBM Analog Hardware Acceleration Kit, and the open source repositories for G\&CNETs and geodesyNets.


\bibliographystyle{elsarticle-num} 
\bibliography{references.bib, library.bib, cas-refs}
\end{document}